\input harvmac
\noblackbox
%\draftmode

\input epsf

\newcount\figno
\figno=0
\def\fig#1#2#3{
\par\begingroup\parindent=0pt\leftskip=1cm\rightskip=1cm\parindent=0pt
\baselineskip=11pt
\global\advance\figno by 1
\midinsert
\epsfxsize=#3
\centerline{\epsfbox{#2}}
\vskip 12pt
{\bf Fig.\ \the\figno: } #1\par
\endinsert\endgroup\par}
\def\figlabel#1{\xdef#1{\the\figno}}
\def\encadremath#1{\vbox{\hrule\hbox{\vrule\kern8pt\vbox{\kern8pt
\hbox{$\displaystyle #1$}\kern8pt}
\kern8pt\vrule}\hrule}}

\def\ph{{\phi}}

\def\cp{{\overline \psi}}

\def\frac#1#2{{#1\over #2}}

%%%%%%%%%%%%%%%%%%%%%%%%%%%%%%%%%%%%%%%%%%%%%%%%%%%
\Title {\vbox{
\baselineskip12pt
\hbox{hep-th/0002186}\hbox{PUPT-1920}\hbox{IASSNS-HEP-00/13}}}
{\vbox{
\centerline{Comments on Noncommutative Perturbative Dynamics}}}

\centerline{Mark Van Raamsdonk\footnote{$^1$}{mav@princeton.edu}}
\smallskip
\centerline{\sl Department of Physics, Princeton University}
\centerline{\sl Princeton, NJ 08544, USA}
\medskip
\centerline{and}
\medskip
\centerline{Nathan Seiberg\footnote{$^2$}{seiberg@sns.ias.edu}}
\smallskip
\centerline{\sl School of Natural Sciences, Institute for Advanced 
Study}
\centerline{\sl Olden Lane, Princeton, NJ 08540, USA}

\vskip 0.6cm

\centerline 
{\bf Abstract} 
\medskip 
\noindent
We analyze further the IR singularities that appear in noncommutative
field theories on $\CR^d$.  We argue that all IR singularities in
nonplanar one loop diagrams may be interpreted as arising from the
tree level exchanges of new light degrees of freedom, one coupling to each relevant operator. These exchanges are reminiscent of
closed string exchanges in the double twist diagrams in open string
theory.  Some of these degrees of freedom are required to have propagators that are inverse linear or logarithmic. We suggest that these can be interpreted as free propagators in one or two extra dimensions respectively.  We also calculate some of the IR singular terms appearing at two loops in noncommutative scalar field theories and find a complicated momentum dependence which is more difficult to interpret.

\vskip 0.5cm
\Date{February 2000}

\newsec{Introduction}

\nref\rfilk{T.~Filk, ``Divergences in a Field Theory on Quantum
Space,'' Phys. Lett. {\bf B376} (1996) 53.}%
\nref\onevgb{J.C.~Varilly and J.M.~Gracia-Bondia, ``On the ultraviolet
behavior of quantum fields over noncommutative manifolds,''
Int.\ J.\ Mod.\ Phys.\ {\bf A14} (1999) 1305, hep-th/9804001.}%
\nref\two{M.~Chaichian, A.~Demichev and P.~Presnajder, ``Quantum Field
Theory on Noncommutative Space-times and the Persistence of
Ultraviolet Divergences,'' hep-th/9812180;  ``Quantum Field Theory on
the Noncommutative Plane with E(q)(2) Symmetry,'' hep-th/9904132.}%
\nref\rjabbari{ M. Sheikh-Jabbari, ``One Loop Renormalizability
of Supersymmetric Yang-Mills Theories on Noncommutative Torus,''
hep-th/9903107, JHEP {\bf 06} (1999) 015.}%
\nref\rruiz{C.P.Martin, D. Sanchez-Ruiz,
``The One-loop UV Divergent Structure of U(1) Yang-Mills
 Theory on Noncommutative $R^4$,''hep-th/9903077,
Phys.Rev.Lett. {\bf 83} (1999) 476-479.}%
\nref\rwulkenhaar{T. Krajewski, R. Wulkenhaar, ``Perturbative quantum
gauge fields on the noncommutative torus,'' hep-th/9903187.}%
\nref\twopfo{S.~Cho, R.~Hinterding, J.~Madore and H.~Steinacker,
``Finite Field Theory on Noncommutative Geometries,''
hep-th/9903239.}%
\nref\three{E.~Hawkins, ``Noncommutative Regularization for the
Practical Man,'' hep-th/9908052.}%
\nref\rsusskind{D. Bigatti and L. Susskind, 
``Magnetic fields, branes and noncommutative geometry,'' 
hep-th/9908056.}%
\nref\rishibashi{N. Ishibashi, S. Iso, H. Kawai and Y. Kitazawa,
``Wilson Loops in Noncommutative Yang-Mills,'' hep-th/9910004.}%
\nref\riouri{I. Chepelev and  R. Roiban, 
``Renormalization of Quantum Field Theories on Noncommutative $R^d$, I. 
Scalars,'' hep-th/9911098.}%
\nref\rbenaoum{H. Benaoum, ``Perturbative BF-Yang-Mills theory 
on noncommutative $R^4$,'' hep-th/9912036.}%
\nref\msv{S. Minwalla, M. Van Raamsdonk, N. Seiberg ``Noncommutative 
Perturbative Dynamics,'' JHEP {\bf 02} (2000) 020, hep-th/9912072.}%
\nref\arcioni{G.~Arcioni and M.~A.~Vazquez-Mozo,
``Thermal effects in perturbative noncommutative gauge theories,''
JHEP {\bf 0001} (2000) 028, hep-th/9912140.}%
\nref\haya{M.~Hayakawa,
``Perturbative analysis on infrared aspects of noncommutative QED on
R**4,'' hep-th/9912094;
``Perturbative analysis on infrared and ultraviolet aspects of  
noncommutative QED on R**4,'' hep-th/9912167.}%
\nref\ikk{S.~Iso, H.~Kawai and Y.~Kitazawa,
``Bi-local fields in noncommutative field theory,'' hep-th/0001027.}%
\nref\grosse{H.~Grosse, T.~Krajewski and R.~Wulkenhaar,
``Renormalization of noncommutative Yang-Mills theories: A simple  
example,''
hep-th/0001182.}%
\nref\arefeva{I.~Y.~Aref'eva, D.~M.~Belov and A.~S.~Koshelev,
``A note on UV/IR for noncommutative complex scalar field,''
hep-th/0001215.}%
\nref\texa{W.~Fischler, E.~Gorbatov, A.~Kashani-Poor, S.~Paban,
P.~Pouliot and J.~Gomis, 
``Evidence for winding states in noncommutative quantum field theory,''
hep-th/0002067.}%
\nref\sunew{A.~Matusis, L.~Susskind and N.~Toumbas,
``The IR/UV connection in the non-commutative gauge theories,''
hep-th/0002075.}%

In this note we continue the analysis of the
perturbation expansion of field theories on noncommutative ${\cal
R}^d$ \refs{\rfilk-\sunew}.  We consider theories in various dimensions
defined by an action
\eqn\gena{S = \int d^d x \; \tr \big( {1 \over 2} (\partial \phi)^2 + {1 \over
2} m^2 \phi^2  + \sum_n {\lambda_n } \overbrace{ \phi \star \phi \star
\cdots \phi }^{n} \big) \; ,} 
where $\star$ is the noncommutative, associative star product defined
by 
\eqn\stard{f \star g (x) = e^{{i \over 2} \Theta^{\mu \nu}
\partial^x_\mu  \partial^y_\nu} f(x) g(y) |_{y=x},}
and $\Theta$ is a constant anticommuting noncommutativity matrix, 
\eqn\noncomt{[x^\mu, x^\nu] = i \Theta^{\mu \nu} \; .}

In \msv , perturbative properties of these noncommutative scalar field 
theories were investigated through the explicit calculation of 
correlation functions. The $\star$-product form of the interactions 
leads to a momentum dependent phase associated with each vertex of a 
Feynman diagram. This phase is sensitive to the order of lines entering 
the vertex, so different orderings 
lead to diagrams with very different behavior. As was first demonstrated 
by Filk \rfilk , planar diagrams (with no crossings of lines) differ 
{}from the corresponding diagrams in the commutative theory only by external momentum dependent phase factors. These graphs lead to single trace terms like \gena\ in the effective action, including divergent terms which renormalize the bare action. The Feynman integrals for these graphs are the same as in the commutative case, resulting in the usual UV divergences which may be dealt with in the usual way by introducing counterterms. 

Nonplanar diagrams contain internal momentum dependent phase factors 
associated with each crossing of lines in the graph. The oscillations of 
these phases serve to lessen any divergence, and may render an otherwise 
divergent graph finite, providing an effective cutoff $\Lambda_{eff} 
= {1 \over \sqrt{\theta}}$ in cases when internal lines cross ($\theta$ 
is a typical eigenvalue of $\Theta^{\mu \nu}$) or $\Lambda_{eff} = {1 
\over \sqrt{-p_\mu (\Theta^2)^{\mu \nu} p_\nu}} \equiv {1 \over \sqrt{p 
\circ p}}$ in cases where an external line with momentum $p$ crosses an 
internal line. In the latter case, we see that the original UV 
divergence is replaced with an IR singularity, since taking $p \to 0$ 
results in $\Lambda_{eff} \to \infty$.

This striking occurrence of IR singularities in massive theories
suggests the presence of new light degrees of freedom. Indeed, an
analysis of the one loop corrected propagator of $\phi$ reveals that
in addition to the original pole at $p^2 \approx - m^2$, there is a new
pole at $p^2 = {\cal O}(g^2)$. This new pole can be understood as
arising from the high momentum modes of $\phi$ running in a loop. If
we try to use a Wilsonian effective action with a fixed cutoff
$\Lambda$, these modes are absent, and indeed, we find that the $p \to
0$ limit is not singular, since the effective cutoff
$\Lambda_{eff}$ is replaced by the cutoff $\Lambda$ when $p \circ p <
1/\Lambda^2$. Thus the $p \to 0$ and $\Lambda \to \infty$ limits do
not commute. In order to write a Wilsonian effective action that does
correctly describe the low momentum behavior of the theory, it is
necessary to introduce new fields into the action which represent the
light degrees of freedom. In \msv , it was shown that the quadratic
IR divergences in the two point functions of $\phi^4$ in four dimensions 
or $\phi^3$ in six dimensions can be reproduced by adding a field 
$\chi_0$ with action of
the form
\eqn\schi{
S_{\chi_0} = \int d^dx \; g  \chi_0 \tr(\phi) + {1 \over 2} \partial \chi_0 
\circ 
\partial \chi_0 + {1 \over 2} \Lambda^2 (\partial \circ \partial 
\chi_0)^2.}
With this action, the quadratic IR singularity in the two point function 
of $\phi$ is reproduced by a diagram in which $\phi$ turns into $\chi_0$ 
and back into $\phi$. 

It should be stressed that in Lorentzian signature spacetime with
$\Theta^{0i} = 0$ the field $\chi_0$ is not dynamical.  It is a
Lagrange multiplier \msv.  Yet, it does lead to long range
correlations.  Even though it is not a propagating field in this case,
we will loosely refer to it as a particle.

These effects are very reminiscent of channel duality in string theory
\msv. There, high momentum open strings running in a loop have a dual
interpretation as the exchange of a light closed string. By this
analogy, we may associate the field $\phi$ with the modes of open
strings, while $\chi_0$ describes a closed string mode.
We thus see that noncommutative field theories are interesting toy
models of open string theories.  Other evidence to the stringy nature
of these theories is their T-duality behavior when they are
compactified on tori and their large $\Theta$ behavior \msv.

Clearly, the appearance of these closed string modes is a generic
phenomenon occurring whenever the commutative theory exhibits UV
divergences.  It is surprising because the zero slope limit of 
\ref\rsw{N. Seiberg and E. Witten, ``String Theory and Noncommutative
Geometry,'' hep-th/9908142, JHEP { \bf 09} (1999) 032.}
is supposed to decouple all the higher open string modes of the string
as well as the closed string modes.  In hindsight this phenomenon is
perhaps somewhat less surprising.  The fields living on a brane are
modes of open strings.  The parameters in this theory are the zero
momentum modes of the closed string background in which the brane is
embedded.  If the theory on the brane is not conformal, the
renormalization group in the theory on the brane changes the values of
these parameters.  Therefore, it is typical for the zero momentum
modes of the closed strings not to decouple.  We now see that in the
noncommutative theories, the nonzero modes also fail to decouple.

Given this understanding of the quadratic IR singularities as poles of a
light particle, it is interesting to ask whether we can find a similar
interpretation for the inverse linear and logarithmic IR singularities
that appear.  These occur wherever linear or logarithmic UV divergences
appear in the commutative theories, including two point functions and
higher point interactions.  It is the goal of this paper to provide some
further understanding of these logarithmic and inverse linear IR
singularities.

In the next section, we determine the complete set of IR singularities
in the low energy effective action that arise from one loop graphs in
various scalar theories.  We show that they may be reproduced by
including a set of $\chi$ fields coupling to each relevant operator of
the theory, if we allow some of the fields $\chi$ to have propagators
which behave like $\ln(1 / p \circ p)$ or $(p \circ p)^{-1/2}$.  In
section 3, we point out that these propagators arise naturally if the
$\chi$ fields are actually free particles in extra dimensions, coupling
to $\phi$s that live on a brane of codimension two for logarithmic
singularities or codimension one for $(p \circ p)^{-1/2}$ singularities.
This is natural given the analogy with string theory, since we associate
the $\phi$ particles with open string modes which live on a brane, while
the $\chi$s are closed strings which should be free to propagate in the
bulk.  In section 4, we consider higher loop graphs in scalar field 
theory, and find IR singularities with more complicated momentum 
dependence that are more difficult to interpret. 

\newsec{Low energy one loop effective action}

In this section, we write down the complete set of IR singular terms in 
the one loop 1PI effective actions of various scalar theories. We will 
take $\phi$ to be an $N \times N$ matrix since it is useful to see how 
the indices are contracted in the various terms. In particular, it turns 
out that all IR singular terms take the form
\eqn\genform{
\tr({\cal O}_1(p)) \tr({\cal O}_2 (-p)) f(p) \; ,
}
where the ${\cal O}$s are operators built out of $\phi$, and $f(p)$ 
diverges quadratically, linearly, or logarithmically as $p \to 0$. We 
show that any such term may be understood as arising from the exchange 
of a single scalar particle which couples separately to $\tr({\cal 
O}_1)$ and $\tr({\cal O}_2)$ and which has a propagator $f(p)$. 

\subsec{$\phi^3$ theory in $d=4$}

We begin by considering the simple example of $\phi^3$ theory in four
dimensions
\eqn\phithreef{S= \int d^4 x \; \tr \left({1 \over 2} (\partial \phi)^2 
+ {1 \over
2} m^2 \phi^2  + {g \over 3!} \phi \star \phi \star \phi \right) \; .}
The commutative theory is superrenormalizable, 
and the only UV divergences are logarithmic divergences in the one loop 
contributions to the 1PI effective action. These come from the planar 
and nonplanar diagrams shown in figure 1 which contribute respectively 
to $\tr(\phi^2)$ and $\tr(\phi)\tr(\phi)$ terms in the effective action. 

\fig{Planar and nonplanar contributions to the one loop quadratic 
effective action in $\phi^3$ theory in four dimensions.} 
{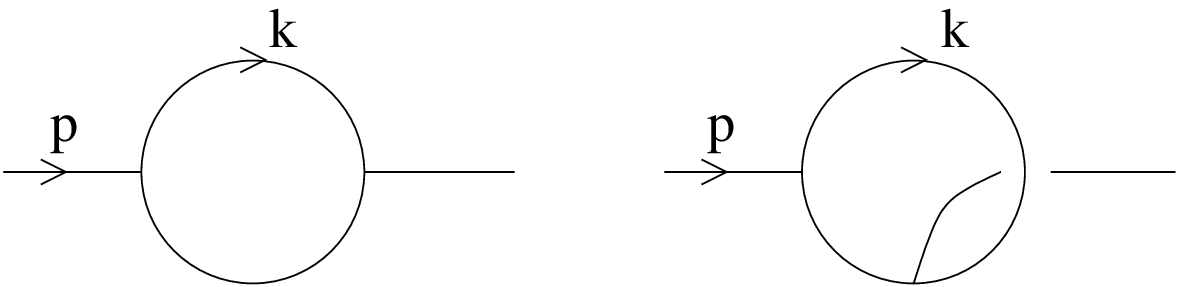}{3.0truein}

In the noncommutative theory, the planar diagram is unchanged, while the 
nonplanar diagram becomes finite, cutoff by $\Lambda^2_{eff}= {1 \over 
p \circ p}$. Combining the two contributions, we find that the one loop 
quadratic effective action (at finite cutoff) is 
\eqn\effact{\eqalign{
S_{eff}=(2 \pi)^4 \int d^4 p  & \; \; {N \over 2} \tr(\ph(p) \ph(-p))  
(p^2 + M^2) \cr
& - {1 \over 2} \tr(\phi(p)) \tr(\phi(-p))
{g^2 \over 64 \pi^2} \ln\left( { 1\over M^2 (p \circ p +1 / 
\Lambda^2 )}\right)
+ ... \;  , \cr}} 
where $M$ is the planar renormalized mass, corrected at one loop by the 
planar diagram in figure 1 plus a counterterm graph. 

The second term in \effact\ arises from the nonplanar diagram and 
contains a logarithmic IR singularity for $\Lambda = \infty$ but not at 
finite cutoff. Thus, as in \msv , the $\Lambda \to \infty$ and $p \to 
0$ limits do not commute. In order to reproduce the correct low momentum 
behavior in a Wilsonian action, we must introduce a new field. 

As for the case of theories with quadratic divergences considered in 
\msv , we introduce a new field $\chi$ which couples to $\tr(\phi)$ 
\eqn\chicoup{
\int d^d x \;  g \chi(x) \tr(\phi(x)) \; .
}
Now, suppose that $\chi$ has a propagator given by
\eqn\chiprop{
\langle \chi(p) \chi(-p)\rangle  = f(p) \; .
}
Then upon integrating out $\chi$, the quadratic effective action for 
$\phi$ receives a contribution
\eqn\neweff{
S = (2 \pi)^d \int d^d p {1 \over 2} \tr(\phi(p))\tr(\phi(-p)) \left( - 
g^2 f(p) \right) \; .}
Thus, we see that the logarithmic singularity in \effact\ may be 
reproduced by including a coupling \chicoup\ in the Wilsonian effective 
action, if $\chi$ has a momentum space propagator
\eqn\lnprop{
f(p) = {1 \over 64 \pi^2} \ln \left({p \circ p + 1/ \Lambda^2 \over p 
\circ p}\right) \; .
}
The numerator in the logarithm has been chosen to cancel the incorrectly 
cutoff logarithm coming from the nonplanar $\phi$ loop in the cutoff 
theory. 

In this theory, there are no additional singularities at higher loops, 
so this single new field is enough to reproduce all IR singularities. 
In section 3, we will give a possible interpretation of this logarithmic 
propagator, but first we turn to more complicated examples. 

\subsec{$\phi^4$ theory in $d=4$}

As a second example, we consider $\phi^4$ theory in four dimensions
\eqn\phiff{S= \int d^4 x \; \tr \left( {1 \over 2} (\partial \phi)^2 + 
{1 \over
2} m^2 \phi^2  + {g^2 \over 4!} \phi \star \phi \star \phi \star \phi 
\right) \;.}
The effective action for the case where $\phi$ is not a
matrix ($N=1$) was computed in \msv\  (for low momenta),
\eqn\effectiveactionb{\eqalign{
S_{eff} & =  (2 \pi)^4 \int d^4 p  {1 \over 2} \ph(p) \ph(-p) \left( p^2 
+ M^2 + {g^2 \over 96 \pi^2 (p \circ p + {1 \over \Lambda^2})} \right. 
\cr &  \left. \qquad \qquad \qquad \qquad \qquad \qquad \qquad -
{g^2 M^2 \over 96 \pi^2} \ln\left( { 1\over M^2( p \circ p + {1 \over 
\Lambda^2})}\right) 
+ ... \right) \cr & + (2 \pi)^4 \int d^4 p_i {1 \over 4!} \phi(p_1) 
\phi(p_2) \phi(p_3) \phi(p_4) \delta(\sum p_i) \cr 
& \qquad \qquad  \left( 
g^2 - {g^4 \over 3 \cdot 2^5 \pi^2}\sum_i \ln \left({1 \over M^2 (p_i 
\circ p_i + {1 \over \Lambda^2})} \right) \right. \cr &  \left. \qquad 
\qquad \qquad  - {g^4 \over 3 \cdot 2^6 \pi^2}\sum_{i < j} \ln 
\left({1 \over M^2 ((p_i + p_j) \circ (p_i + p_j) + {1 \over 
\Lambda^2})} \right) + ...\right). 
\cr}}

In this case, the quadratic effective action has both quadratic and 
logarithmic IR singularities for $\Lambda = \infty$, while the quartic 
term has two types of logarithmic singularities. In \msv , it was shown 
that the ${1 \over p \circ p}$ term in the $\Lambda \to \infty$ 
quadratic effective action is reproduced by a Wilsonian effective 
action which includes a $\chi_0$ field coupling to $\phi$ with action of the form \schi .

We now focus on the terms containing logarithmic singularities. It is 
illustrative to generalize to the case of arbitrary $N$ and write them
as
\eqn\split{\eqalign{
& \int d^4 x d^4 y \left\{ -g^2 M^2 \tr(\phi(x)) \tr( \phi(y)) - 
{g^4 \over 3} \tr(\phi(x)) \tr( \phi^3(y)) -{g^4 \over 
4} \tr(\phi^2(x))\tr(\phi^2(y)) \right\} \cr
& \qquad \qquad \qquad \qquad \cdot {1 \over 3 \cdot 2^6 \pi^2} \int {d^4p 
\over (2 \pi)^4} e^{i p \cdot (x-y)} 
\ln \left( {1 \over M^2 (p \circ p + {1 \over \Lambda^2})} \right) \cr
& = \int d^4x d^4y \sum g^{m+n} M^{4-m-n} \gamma_{mn} \tr(\phi^m(x)) 
\tr(\phi^n(y)) \Delta(x-y)  \; ,\cr}}
where
\eqn\deltaxy{
\Delta(x-y) = \int {d^4p \over (2 \pi)^4} e^{i p \cdot (x-y)} \ln \left( 
{1 \over M^2 (p \circ p + {1 \over \Lambda^2})} \right),}
and $\gamma_{mn}$ are numerical constants that may be read off from 
\split.
In this form it is clear that the complete set of logarithmic IR 
singularities in the $\Lambda \to \infty$ one loop effective action may 
be reproduced with a finite cutoff Wilsonian action by 
including $\chi$ fields with couplings 
\eqn\indfxsu{\int d^4x \sum_{n=1}^3 g^n M^{2-n} \chi_n(x)
\tr(\phi^n(x))} 
and logarithmic propagators
\eqn\logpro{
\langle \chi_m(p) \chi_n(-p)\rangle  = -2 \gamma_{mn} \ln({p \circ p + 
{1 \over \Lambda^2} \over p 
\circ p}) \; .}
In this way, the $\Lambda=\infty$ IR singularity in each term of \split\
is reproduced at finite cutoff by the exchange of a single scalar 
particle, as shown in figure 2.

\fig{Examples of nonplanar diagrams in $\phi^4$ in $d=4$ contributing
IR singularities for $\Lambda \to \infty$ and $\chi$ exchange diagrams
that reproduce the singularities at finite
cutoff.}{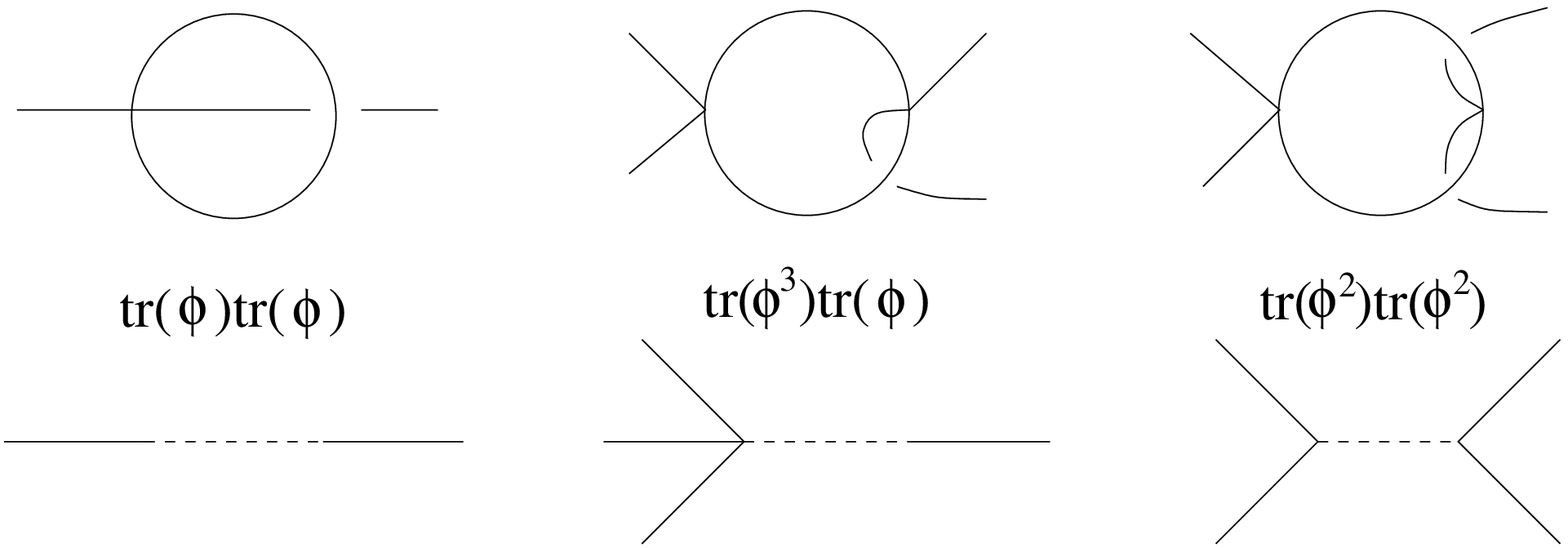}{4.0truein}

\subsec{$\phi^3$ theory in $d=6$}

As a final example, we consider $\phi^3$ theory in six dimensions,
\eqn\phics{S=  \int d^6 x \; \tr \left( {1 \over 2} (\partial \phi)^2 + 
{1 \over
2} m^2 \phi^2  + {g \over 3!} \phi \star \phi \star \phi \right) \;.}
Here, the one loop effective action (for low momenta) was computed in
\msv\ for $N=1$,
\eqn\efff{\eqalign{
S_{eff}  & =  (2 \pi)^6 \int d^6 p {1 \over 2} \ph(p) \ph(-p) 
\left( p^2+ M^2 - {g^2 \over 2^8 \pi^3 (p \circ p + {1 \over \Lambda^2}) 
} \right. \cr & \qquad \qquad \qquad \qquad \qquad \left. + {g^2 \over 3 
\cdot 2^9  \pi^3}(p^2 +6M^2) \ln({1 \over M^2 ( p \circ p + {1 \over 
\Lambda^2}) })+... \right) \cr
& + (2 \pi)^6 \int d^6 p_i {1 \over 3!} \phi(p_1) \phi(p_2) \phi(p_3) 
\delta(\sum 
p_i) \left\{ g + {g^3 \over 2^9 \pi^3}\sum_i \ln \left( {1 \over M^2 
(p_i \circ p_i + {1 \over \Lambda^2})} \right) \right\} \; . \cr}}
The IR singularities in this action at $\Lambda = \infty$ are similar to 
those of the $\phi^4$ theory, but now we have a $p^2 \ln({1 \over M^2 p 
\circ p })$ term in the quadratic effective action. We may rewrite the 
terms with logarithmic singularities for arbitrary $N$ as 
\eqn\splita{\eqalign{
\int d^6 x d^6 y & \left\{ gM \; \tr(\phi(x)) \left( g M  
\tr(\phi(y)) - {g \over 6 M} \tr(\partial^2 
\phi(y)) +{ g^2 \over 2 M} \tr( \phi^2(y) ) \right) \right\} \cr & \qquad \qquad \cdot
{1 \over 512 \pi^3} \int {d^6p \over (2 \pi)^6} e^{i p \cdot (x-y)} \ln \left( {1 \over M^2 p \circ p} 
\right) \; . \cr }}
These may be reproduced by introducing $\chi$ fields with couplings
\eqn\cp{
\int d^6x g M \; \chi_1(x) \; \tr(\phi(x)) +  \chi_2 (x) \left( g M \tr( 
\phi(x)) + 
{g^2 \over 2M} \tr(\phi^2(x)) -   {g \over 6M} \tr( \partial^2 \phi(x)) 
\right)}
and propagators
\eqn\chiprop{\eqalign{
\langle \chi_1(p) \chi_2(-p)\rangle  & =  -{1 \over 512 \pi^3 } \ln 
\left({p \circ p + 1 / 
\Lambda^2  \over p\circ 
p} \right) \cr
\langle \chi_1(p) \chi_1(-p)\rangle  & =\langle \chi_2(p) 
\chi_2(-p)\rangle  = 0. \cr}}
By a linear redefinition of the $\chi$s, we may simplify the couplings
to 
\eqn\actionofchi{
\int d^6x  \; g M \hat{\chi}_1(x) \tr(\phi(x)) +  \hat{\chi}_2 (x) 
\left( {g^2 
\over 2M}  \tr(\phi^2(x)) - {g \over 6M} \tr(\partial^2 \phi) \right),}
where $\hat{\chi}_1 = \chi_1 + \chi_2$ and $\hat{\chi_2} = \chi_2$.

\subsec{General procedure}

The three examples we have considered lead us to a general procedure for 
introducing $\chi$ fields in the Wilsonian effective action in order to 
reproduce logarithmic singularities in the one loop effective action. 
For a general scalar theory, logarithmic IR singular terms in the 
effective action arising from one loop non-planar diagrams take the 
form
\eqn\gen{\sum_{m,n} \int d^4x d^4y {1 \over 2} {\cal O} _m(\phi(x)) 
{\cal O}
_n(\phi(y)) \gamma_{mn} \int {d^dp 
\over (2 \pi)^d }e^{i p\cdot (x-y)} \ln\left({1 \over m^2 (p \circ p +
{1  \over \Lambda^2})}\right).}
Here $\{{\cal O} _n(\phi(x))\}$ is some basis for the set of relevant
local operators (such as $\tr(\phi^m)$, $\tr(\partial^2 \phi)$, etc...).
$\gamma_{mn}$ is a ``metric'' on the space of operators, which we may
take to be a matrix of numerical constants by assuming that all masses
and coupling constants are included in the ${\cal O} $s. Note that
terms in the effective action at higher loops will involve products of 
more than
two operators, however the one loop terms may always be written in
this form. We now introduce a $\chi$ field coupling to each ${\cal O}$,
\eqn\schiphi{S_{\chi \phi} = \int d^dx \chi_n(x) {\cal O} _n(\phi(x)),}
and assume that the fields $\chi$ have propagators
\eqn\chiprop{\langle \chi_m(p) \chi_n(-p)\rangle  = -\gamma_{mn} \ln
\left({p \circ p +  {1 \over \Lambda^2} \over p \circ p}\right)}
which could arise, for example, from the nonlocal quadratic action
\eqn\nloc{
S_{\chi \chi} = -\int d^dp {1 \over 2} \chi_m(p) \chi_n(-p) \gamma^{mn}
\left[\ln({p \circ p + {1 \over \Lambda^2} \over p \circ p})
\right]^{-1}.} 
Here, $\gamma^{mn}$ is the inverse \foot{If $\gamma$ is not
invertible, we have simply introduced too many $\chi$s. In this case,
we choose a new basis $\{\hat{{\cal O} }_n\}$ of operators such
that some of the basis elements do not appear in \gen .  The
submatrix $\hat{\gamma}$ of $\gamma$ corresponding to the $\hat{{\cal
O}}$s which do appear will then be invertible, and we may introduce
$\chi$s as above coupling to this smaller set of operators. The
kinetic term \nloc\ for the $\chi$s is then well defined, since
we replace $\gamma$ with $\hat{\gamma}$.} of $\gamma_{mn}$. 

\subsec{Linear divergences in $\phi^4$ in three dimensions}

Before closing this section, we note that linear IR singularities at one 
loop may be understood in a similar manner. For example, the commutative 
$\phi^4$ theory in $d=3$ has a linear divergence in the two point 
function at one loop. In the noncommutative theory, this leads to a term 
\eqn\linac{
S_{IR} = \int d^3 p \;  {1 \over 2} \tr(\phi(p)) \tr(\phi(-p)) {\pi^2 
\over 6 (p \circ p + {1 \over \Lambda^2})^{1 \over 2} }
}
in the quadratic effective action which has a $1/|p|$ singularity for
$\Lambda = \infty$. In order to reproduce this singularity, we again 
introduce a $\chi$ field coupling to $\tr(\phi)$ but this time we need a 
propagator
\eqn\linprop{
\langle \chi(p) \chi(-p)\rangle  \propto {1 \over (p \circ p)^{1 \over 
2}}  - {1 \over (p \circ p + {1 \over \Lambda^2})^{1 \over 2}} \; .
}

\newsec{What can give logarithmic and inverse linear propagators?}

In the previous section we showed that the singular IR behavior of the 
one loop effective actions for various scalar field theories can be 
reproduced by a Wilsonian action which includes new particles $\chi_i$ 
coupling linearly to various operators built from of $\phi$, assuming 
that the fields $\chi$ have logarithmic (or inverse linear) propagators. 
We would now like to understand what dynamics could give rise to these 
propagators. 

\subsec{Spectral representation of propagators}

As a first step, it is useful to rewrite the propagators in a spectral
representation. We introduce a parameter\foot{We call the constant
$\alpha'$ since for noncommutative field theories arising from string
theory, the metric $g^{\mu\nu}=-{(\Theta^2)^{\mu \nu} \over
(\alpha')^2}$ is exactly the closed string metric in the zero slope
limit of \rsw\ (the open string metric is $G_{\mu \nu} =
\delta_{\mu \nu}$) when $\alpha'$ has its usual meaning.}  $\alpha'$ 
with dimensions of squared length and a metric 
$g^{\mu\nu}=-{(\Theta^2)^{\mu \nu} \over  (\alpha')^2}$ then rewrite the 
propagator as
\eqn\deltaref{
\Delta(p) = \int d m^2 \rho(m^2) {1 \over {p \circ p \over (\alpha')^2} 
+ m^2} \; .}

The logarithmic propagators
\eqn\logprop{
\Delta(p) = \ln\left( { p \circ p + {1 \over \Lambda^2} \over p \circ p} 
\right)}
are reproduced with the spectral density
\eqn\speca{
\rho(m^2) = \left\{ \matrix{ 1 & m^2 \le {1 \over ( \alpha' \Lambda)^2} 
\cr 0 & m^2 > {1 \over (\alpha' \Lambda)^2} \cr} \right.}
suggesting a continuum of states with $m^2$ uniformly distributed 
between 0 and a cutoff $1/(\alpha' \Lambda)^2$. Thus, we may replace 
$\chi$ with a continuum of states $\psi_m$ which have ordinary 
propagators
\eqn\psiopro{
\langle \psi_m(p) \psi_m(-p) \rangle \propto {1 \over {p \circ p 
\over(\alpha')^2 } + m^2} }
which couple to $\phi$ (or some ${\cal O}(\phi)$) in a way that is 
independent of $m$,
\eqn\spsiphi{
S_{\psi \phi} = \int^{1 \over (\alpha' \Lambda)^2} d m^2 \int d^d x 
\psi_m(x) \phi(x).}
Note that for $\Lambda \to \infty$, the $\psi$s completely decouple, 
leaving the original $\phi$ theory, as desired. 

The $\Delta(p)= {1 \over (p \circ p)^{1 \over 2} }$ propagators, are
reproduced by a spectral 
density\foot{The symbols $\roughly{>}$ and $\roughly{<}$ are used 
because the second term in \linprop\ is regularization dependent and its 
precise functional form is not reproduced by choosing a sharp cutoff 
here, though the behavior for $p \to 0$ at fixed $\Lambda$ and for $\Lambda \to \infty$ is the same.}    
\eqn\specb{
\rho(m^2) = \left\{ \matrix{ {1 \over m \pi \alpha'} & m^2 \roughly{<}
{4 \over  (\Lambda \pi \alpha')^2} \cr 0 & 
m^2 \roughly{>} {4 \over (\Lambda \pi \alpha')^2}. \cr} \right.}
As above, we may interpret this as a continuum of states $\psi_m$ 
coupling to $\phi$, this time with density $1/(\alpha' m)$ up to a 
cutoff $m^2 \approx {1 \over (\alpha' \Lambda)^2}$.   

\subsec{$\chi$s as degrees of freedom in extra dimensions} 

A very simple possibility for the interpretation of the continuum of
degrees of freedom $\psi_m$ is that they are the transverse momentum
modes of a particle $\psi$ which propagates freely in more dimensions.
The continuous parameter $m$ is related to the momentum $q$ in these
new dimensions through $m^2=-q^2$.  That is, we imagine that the
$d$-dimensional space in which the $\phi$ quanta propagate is a flat
$d$-dimensional brane residing in a $d+n$ dimensional space. The
$\psi$ particles propagate freely in this space but couple to the
$\phi$ particles on the brane (located at $x_\perp = 0$). We choose the 
metric seen by the $\psi$
particles to be $g^{\mu\nu}=-{(\Theta^2)^{\mu \nu} \over (\alpha')^2}$
in the brane directions and $\delta^{\mu\nu}$ in the transverse
directions.

By choosing the number of extra dimensions to be one or two, we can
precisely reproduce the spectral densities \specb\ or \speca\
corresponding to inverse linear or logarithmic propagators
respectively.

Explicitly, with two extra dimensions we have
\eqn\psiprop{\eqalign{
\langle \psi(p,x_\perp \! \!=  \!0) \psi(-p,x_\perp \!\!= \!0) \rangle &= \int^{1 \over
\alpha' \Lambda}  
{d^2 q \over (2 \pi)^2} {1 \over {p \circ p \over (\alpha')^2} + q^2}
= \int^{1 \over (\alpha' \Lambda)^2} {d m^2 \over 4 \pi} {1 \over {p \circ p \over (\alpha')^2} +  m^2} \cr & = {1 \over 4 \pi}  \ln \left( { p \circ p + {1 \over \Lambda^2} \over p \circ p} \right) \; , \cr}}
giving exactly the desired form of the logarithmic propagators. Note 
that we impose a cutoff $1/(\Lambda \alpha')^2$ on $q^2$ to reproduce 
the cutoff in the spectral function \specb . This cutoff on transverse 
momenta {\it decreases} to zero as $\Lambda \to \infty$, but this 
provides the desired decoupling of $\psi$ in this limit.  

With one extra dimension, we find
\eqn\psipropo{
\eqalign{ \langle \psi(p,x_\perp \!\!= \! 0) \psi(-p,x_\perp \! \!= \! 0) \rangle &= \int^{2 \over 
\Lambda \pi  \alpha'} {dq \over 2 \pi} {1 \over {p 
\circ p \over (\alpha')^2} + q^2} = \int^{4 \over (\Lambda \pi 
\alpha')^2} d m^2 {1 \over 2 \pi m} {1 
\over {p \circ p \over (\alpha')^2} +m^2} \cr & = {\alpha' \over 2 (p 
\circ p)^{1 \over 2}} - {\alpha' \over \pi (p \circ p)^{1 \over 2}}  
\tan^{-1}\left( {\Lambda \pi (p \circ p)^{1 \over 2} \over 2} \right) 
\cr}}
thus giving a spectral density of $1/m$ and reproducing the correct $p 
\to 0$ behavior of the inverse linear propagators \linprop\ above (as 
discussed in footnote 3, the difference in functional form between the 
second terms here and in \linprop\ is a consequence of the choice of 
regularization scheme). 

Actually, it is possible to see more directly that the theory with free 
$\psi$ particles in two extra dimensions coupling linearly to $\phi$s 
on the brane is precisely equivalent to a theory with particles $\chi$ 
that live on the brane and have logarithmic propagators. Noting that it 
is $\psi(x,x_\perp = 0)$ that couples to $\phi$, we define $\chi(x) = 
\psi(x,x_\perp = 0)$ 
and rewrite the $\psi$ action using a Lagrange multiplier $\lambda(x)$ 
as
\eqn\eqnwithlag{\eqalign{
&\; \; e^{-\int d^d x \; \phi(x) \psi(x,x_\perp = 0)  - \int d^d x d^2 
x_\perp \; {1 \over 2} (\partial \psi)^2  }  
\cr
&= \int[d\lambda][d \chi] e^{-\int d^d x \; \left[ \phi(x) 
\chi(x) + i\lambda(x)[\chi(x) - \psi(x,x_\perp = 0)] \right]  -\int d^d x d^2 x_\perp \;{1 \over 2} (\partial \psi)^2 } \cr 
&= \int[d\lambda][d \chi] e^{-\int d^dp \;\left[ \phi(p) \chi(-p) +  
i\lambda(-p) \chi(p) \right] - \int d^dp d^2q \; \left[ {1 \over 2} 
\psi(-p,-q)(p 
\circ p + q^2) \psi(p,q) - i\lambda(-p) \psi(p,q)) \right] }.
\cr }}
We may then integrate out $\psi$ directly from this action, leaving
\eqn\intoutchi{
e^{-\int d^dp \; \left[ \phi(p) \chi(-p) +  i\lambda(p) \chi(-p) + {1  
\over 2} \lambda(p) \lambda(-p) \int {d^2q \over p \circ p + q^2}  
\right] .}}
Finally, integrating out the Lagrange multiplier $\lambda$ gives
\eqn\intoutlag{e^{-\int d^dp \;\left[ \phi(p) \chi(-p) + {1 \over 
2}\chi(-p)
\chi(p)  \ln^{-1}\left({p\circ p + {1 \over \Lambda^2} \over p \circ
p}\right) \right]}}
as desired.

In the general case, we find that the nonlocal quadratic action \nloc\
derived above may be replaced by an ordinary, local higher dimensional
kinetic term
\eqn\spsipsi{
\hat{S}_{\psi \psi} = -{1 \over 8 \pi} \int d^{d+2}x {1 \over 2} g^{\mu 
\nu} \partial_\mu \psi_m \partial_\nu \psi_n \gamma^{mn},}
where $g^{\mu \nu}$ is taken to be $-{(\Theta^2)^{\mu \nu} \over 
(\alpha')^2}$ in the 
original $d$ directions and $\delta^{\mu\nu}$ in the transverse
directions.  

As an example, the logarithmic terms in the $\phi^3$ theory in six
dimensions may be reproduced using
\eqn\sphits{\eqalign{
\int d^6x &{gm \over 3 \cdot 32 \pi} \phi(x) \psi(x,x_\perp \! \!= \!0) 
+ \left({3g^2 \over 16 \pi m}\phi^2(x) - {g \over 192 \pi m} \partial^2 
\phi(x)\right) \hat{\psi}(x,x_\perp \! \! = \! 0) \cr
& + \int d^8x {1 \over 2} g^{\mu\nu} \partial_\mu \hat{\psi}
\partial_\nu \hat{\psi} - {1 \over 12} g^{\mu\nu} \partial_\mu \psi
\partial_\nu \hat{\psi} .\cr}} 

\subsec{String theory analogy}

The suggestion that certain high momentum degrees of freedom in $\phi$
are dual to fields which propagate in extra dimensions fits rather
nicely with the analogy that associates $\phi$s with open strings and
$\chi$s or $\psi$s with closed strings. In noncommutative gauge
theories that arise as limits of string theory, the fields ($\phi$s)
in terms of which the theory is defined are modes of open strings
living on a D-brane.  In this case, there is a physical bulk in which
closed strings propagate, and the low energy closed string modes are
related to high energy modes of open strings by channel duality. In
particular, a nonplanar one loop diagram of the type we have studied
is topologically equivalent to a string diagram in which a number of
open strings become a closed string which then turns back into open
strings.  The regime in which the open string loop has very high
momenta may be viewed equivalently as the exchange of a very low
momentum closed string, free to propagate in the bulk. This fits very
well with our interpretation that the IR singularities of nonplanar
one loop diagrams should be reproduced by tree level exchanges of a
particle $\psi$ (see figure 3). The connection between $\psi$ and
closed strings is further strengthened by the fact that the $\psi$s do
not carry any matrix indices, and as noted above, the $\psi$s
propagate in a metric which is precisely the closed string metric
identified in \rsw.

\fig{Channel duality in string theory provides a natural understanding 
of the correspondence between nonplanar one loop diagrams and tree level 
$\chi$ exchange diagrams. } {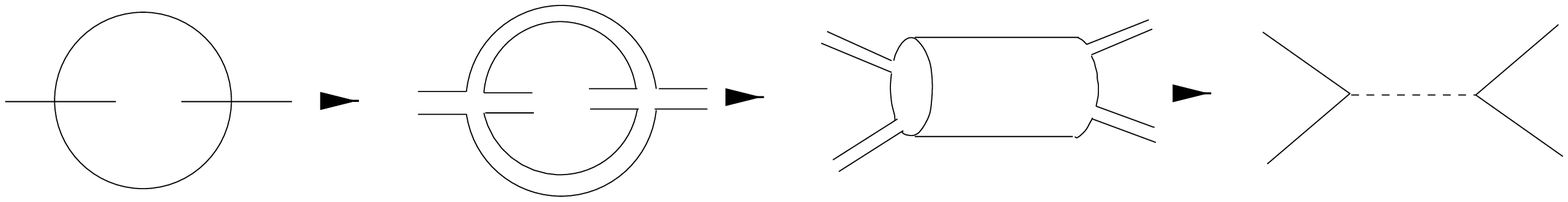}{6.0truein}

\subsec{Interpretations of the extra dimensions}

Despite the many similarities between the $\psi$ fields and closed 
strings, the calculations we have presented so far do not really 
indicate whether the extra dimensions in which the $\psi$s propagate 
are truly physical. There seem to be three logical possibilities:

\item{1)} The first possibility is that the ``extra dimensions'' are
simply a mathematical convenience - a simple way to state that $\chi$
has a logarithmic propagator.

\item{2)} At the other extreme is the possibility that the extra
dimensions are real and the $\psi$s are propagating fields living in
these dimensions whose restrictions to the $d$ dimensional space are
the $\chi$ fields.

\item{3)} The third possibility is a compromise between 
these two: in situations where the noncommutative field theory is a 
limit of string theory, the extra dimensions are really the bulk 
dimensions transverse to the brane on which the $\phi$ fields live and 
the extra dimensions are physical.  Otherwise, they are only a
mathematical convenience.

\noindent
We note that in the case of noncommutative field theories arising from string theory, the number of dimensions transverse to the brane is typically larger than two, so the propagator of a single massless bulk field between points on the brane is nonsingular. However, in these theories there is no reason to expect that only a single closed string mode $\psi$ contributes. Rather, the spectral function required to reproduce IR singularities would presumably arise from the combination of extra dimensions and the density of closed string states which are allowed to couple to the worldvolume field of interest. \foot{We thank Lenny Susskind and Igor Klebanov for helpful discussions on this point.}

\newsec{Higher Loops}

\subsec{The three point function of $\phi^3$ at two loops}

In this section, we explore the infrared singularities appearing in 
$\Gamma^{(3)}$ at two loops for the noncommutative $\phi^3$ theory in
six dimensions. In particular, we consider the two diagrams of figure 4 
which give the leading contribution to the $(\tr\phi)^3$ term in the 
effective action. 

\fig{The two contributions to the $(\tr\phi)^3$ term in the effective 
action at two loops.} {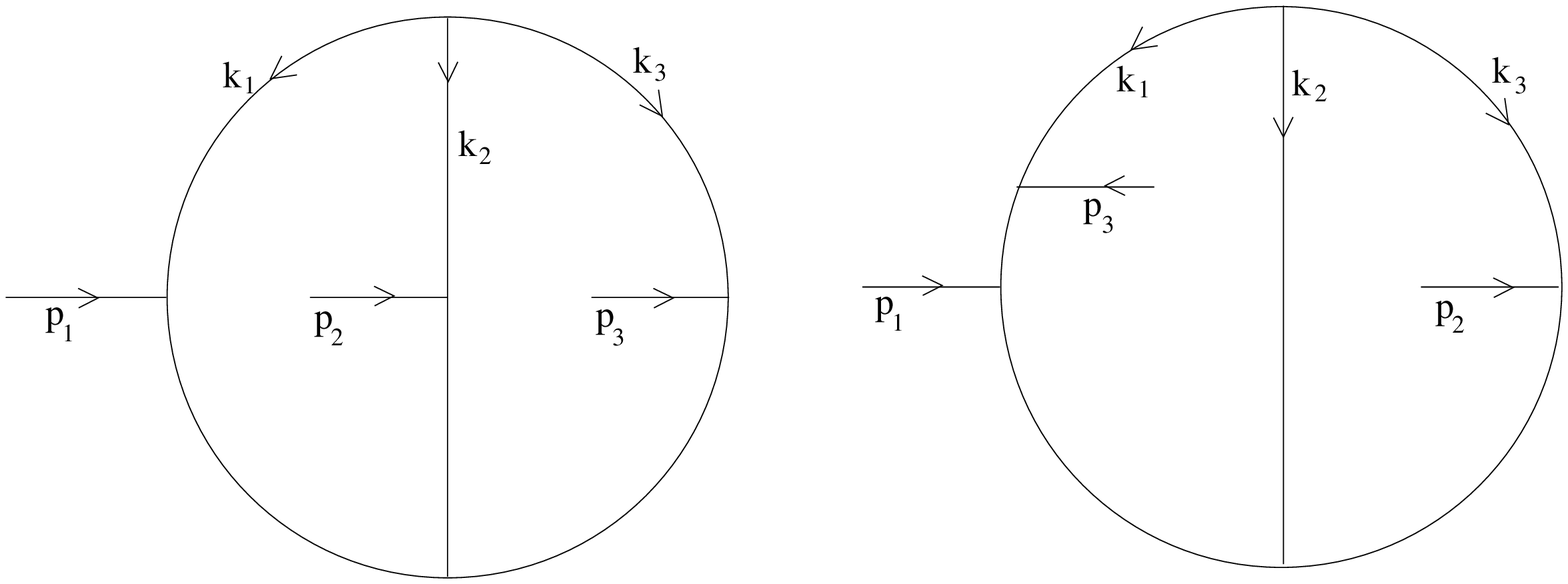}{4.0truein}

These diagrams have the property that each external line connects to a 
different index loop in double line notation. The remaining two loop 
diagrams with three external lines give subleading contributions either 
to $\tr(\phi)tr(\phi^2)$ or to $\tr(\phi^3)$ terms. 

The first diagram contributes a term to the effective action
\eqn\Vone{
\int d^6 p_1 d^6 p_2 d^6 p_3 \delta(p_1 +p_2 +p_3) \phi(p_1) \phi(p_2) 
\phi(p_3) V_1(p_1, p_2, p_3) \; ,
}
where
\eqn\sym{\eqalign{
& V_1(p_1,p_2,p_3) = {g^5 \over 3 \cdot 2^{11} \pi^6} \times \cr & \; \; 
\; \int {d^6 k_1 d^6 k_2 
d^6 k_3 \delta(k_1 +k_2 + k_3)e^{ik_1 \times p_2 - i k_3 \times p_3} 
\over (k_1^2 +m^2)((k_1+p_1)^2 +m^2)(k_2^2 +m^2)((k_2+p_2)^2 +m^2)(k_3^2 
+m^2)((k_3+p_3)^2 +m^2)} \; , \cr }
}
and $k \times p \equiv k_\mu \theta^{\mu \nu} p_\nu$.
The external momenta appearing in the denominator do not contribute to 
the infrared singular terms, since by Taylor expanding the denominators 
in momenta, we find that the subleading terms are nonsingular for $p \to 
0$. To evaluate the leading terms, we rewrite the delta function as 
$\int e^{iy\cdot(k_1 +k_2 +k_3)}$ to obtain
\eqn\syma{
V_1(p_1,p_2,p_3) = {g^5 \over 3 \cdot 2^{11} \pi^6}\int {d^6 y \over (2 
\pi)^6} \prod_{i=1}^3 \int {d^6 k_i e^{i k_i \cdot x_i} \over (k_i^2 + 
m^2)^2} \; ,}
where $x_1 = y + \theta p_2, x_2 = y, x_3 = y - \theta p_3$. The IR 
singular terms come from the region of small $y$, so we need the 
behavior of the $k$ integrals for small $x$. 

In this regime, we have 
\eqn\posprop{
\int {d^6 k_i e^{i k_i \cdot x_i} \over (k^2 + m^2)^2} = {4 \pi^3 \over 
x^2} - m^2 \pi^3 \ln({1 \over x^2 m^2}) + \dots \; .
}
Only the leading term contributes to the IR singular pieces, so our 
Feynman integral becomes
\eqn\symb{
V_1(p_1,p_2,p_3) = {g^5 \pi^3 \over 3 \cdot 2^5} \int^\Lambda {d^6 y 
\over (2 \pi)^6} {1 \over y^2 (y + \theta p_2)^2 (y- \theta p_3)^2} \; 
.}
The cutoff $\Lambda$ has been included because our approximations are 
valid only for small $y$, but this is the only part of the integral that 
gives the IR singular terms of interest. This integral clearly has a 
logarithmic singularity as all momenta go to zero, but is finite, if
any  one momentum is scaled to zero. By a careful analysis of the
integral \symb , it may be shown that for small momenta, we have
\eqn\vonesing{
V_1(p_1,p_2,p_3) = {g^5 \over 3 \cdot 2^{12}} \ln \left({1 \over p_1 
\circ p_1 
+ p_2 \circ p_2 + p_3 \circ p_3 } \right) + {\rm regular \; \; terms} .
}

We now turn to the second diagram of figure 3. This contributes a term 
to the effective action
\eqn\pthree{
\int d^6 p_1 d^6 p_2 d^6 p_3 \delta(p_1 +p_2 +p_3) \phi(p_1) \phi(p_2) 
\phi(p_3) V_2(p_1, p_2, p_3) \; ,
}
where
\eqn\asym{\eqalign{
V_2(p_1,p_2,p_3) = {g^5 \over 2^{10} \pi^6} \int & {d^6 k_1 d^6 k_2 d^6 
k_3 \delta(k_1 +k_2 + k_3)e^{ik_3 \times p_2 - i k_1 \times p_3} \over 
(k_3^2 +m^2)((k_3+p_2)^2 +m^2)(k_1^2 +m^2)((k_1+p_3)^2 +m^2)} \cr
& \qquad \cdot {1 \over ((k_1+p_3 + 
p_1)^2 +m^2)(k_2^2 +m^2)} \; . \cr} 
}
As above, the momenta in the denominator do not affect the IR singular 
terms and we may rewrite the integral as 
\eqn\asyma{
V_2(p_1,p_2,p_3) = {g^5 \over 2^{10} \pi^6} \int {d^6 y \over (2 
\pi)^6}\int {d^6 k_1 d^6 k_2 d^6 k_3 e^{ik_i\cdot x_i} \over (k_1^2 
+m^2)^3(k_2^2 +m^2)(k_3^2 +m^2)^2} \; ,
}
with $x_1 = y - \theta p_3, x_2 = y, x_3 = y + \theta p_2$. The IR 
singular terms come 
only from the region of small $y$. For small $x$, we have
\eqn\propb{\eqalign{
\int {d^6 k_i e^{i k_i \cdot x_i} \over (k^2 + m^2)} & = {16 \pi^3 
\over x^4} - {4 \pi^3 m^2 \over x^2} + {1 \over 2} m^4 \pi^3 \ln({1 
\over 
x^2 m^2}) + \dots \cr
\int {d^6 k_i e^{i k_i \cdot x_i} \over (k^2 + m^2)^2} & = {4 \pi^3 
\over x^2} - m^2 \pi^3 \ln({1 \over x^2 m^2}) + \dots \cr
\int {d^6 k_i e^{i k_i \cdot x_i} \over (k^2 + m^2)^4} & = {1 \over 2} 
\pi^3 \ln({1 \over x^2 m^2}) + \dots  \; .\cr}
} 
The IR singular terms come from taking the leading term in these three 
expressions, and we find
\eqn\asyma{
V_2(p_1,p_2,p_3) = {g^5 \pi^3 \over 2^5} \int^\Lambda {d^6 y \over (2 
\pi)^6}{1 \over y^4 (y + \theta p_2)^2}\ln \left({1 \over (y - \theta 
p_3)^2}\right) \; .}
The small momentum behavior of this integral when any one or all three 
momenta are scaled to zero is  reproduced by the function
\eqn\singb{
\hat{V}_2(p_1, p_2, p_3) = {g^5 \over 2^{11}}({1 \over 2}\ln(p_2 \circ 
p_2) \ln(p_2 \circ p_2 + p_3 \circ p_3) - {1 \over 4} \ln^2(p_2 \circ 
p_2 
+ p_3 \circ p_3)- {1 \over 4} \ln(p_2 \circ p_2 + p_3 \circ p_3)) \; .
}
Note that the logarithmic divergence that arises as $p_2 \to 0$ comes 
{}from the three-point subdiagram which would be divergent in the 
commutative theory. On the other hand, setting $p_1$ or $p_3$ to zero 
does not lead to a divergence. It is possible that there is some other 
function with the same singularities as \singb\ whose form would be more 
suggestive of an interpretation by $\chi$s.

For $\phi^3$ theory with a single $\phi$, we should sum the contribution 
of $V_1$ with that of $V_2$ (which is automatically symmetrized in 
momenta when inserted into the expression \pthree ) to get the complete 
two loop contribution to the $(\tr\phi)^3$ term in the effective 
action. In writing a Wilsonian action to reproduce the IR singularities 
of the $\phi^3$ theory, we do not necessarily need to match the behavior 
diagram by diagram; however it is possible to take a theory which 
isolates the contribution $V_1$, for example. In the theory with 
Lagrangian density
\eqn\manyphi{\eqalign{
\tr & \big( {1 \over 2} \sum_i (\partial \phi_i)^2 + {1 \over 2} m^2 
\sum_i \phi_i^2  \cr & \qquad  + g(\phi_1 \star 
\phi_4 \star \phi_4 + \phi_2 \star \phi_5 \star \phi_5 + \phi_3 \star 
\phi_6 \star \phi_6 + \phi_4 \star \phi_5 \star \phi_6 + {\rm c.c.} ) 
\big)\; , \cr}
}
the leading $\tr(\phi_1) \tr(\phi_2) \tr(\phi_3)$ term in the effective 
action is precisely $V_1$. Thus, we would like to understand how the 
singularity $\ln(p_1^2 +p_2^2 +p_3^2)$ alone or terms of the form 
\singb\ could arise from a Wilsonian effective action. 

\subsec{Interpretation of the singularities}

The two loop IR singularities we have found have momentum 
dependence that cannot be reproduced by tree level diagrams with local 
couplings only on the brane (such as the tree level $\chi$ exchange 
diagrams which reproduced all one loop singularities). Each propagator 
and vertex factor of such a diagram is a function of a single sum of 
external momenta, while the singularities we have found, \vonesing\ and 
\singb , cannot be reproduced by a product of such functions.

\fig{Possibilities for the interpretation of two loop IR 
singularities.}{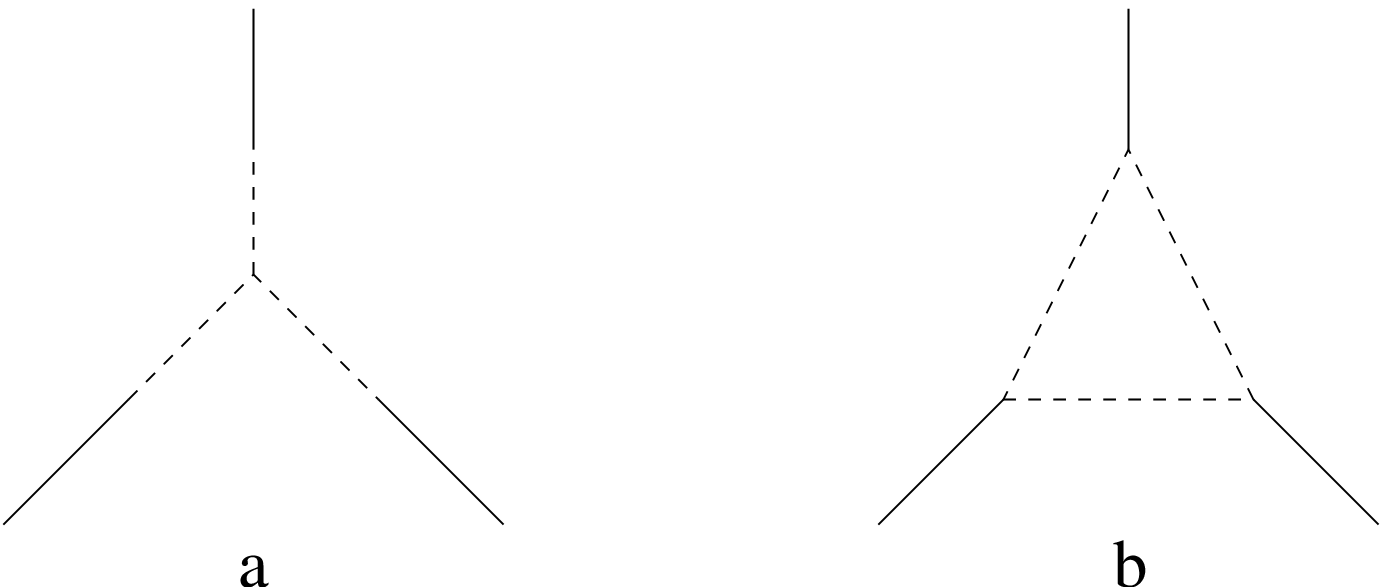}{3.0truein}

If our interpretation of the $\psi$ fields as higher dimensional 
particles is correct, one possibility would be that these singularities 
arise from a diagram like that of Figure 5a, in which each $\phi$ 
becomes a $\psi$ and these three $\psi$ field interact locally in the 
bulk. This is suggested both by the $(\tr\phi)^3$ structure and by the 
fact that viewed as worldsheet diagrams in string theory, the double 
line versions of the diagrams in figure 4 are topologically equivalent 
to the closed string interaction diagram in figure 6.

\fig{Closed string interaction diagram topologically equivalent to the 
diagrams of figure 4.}{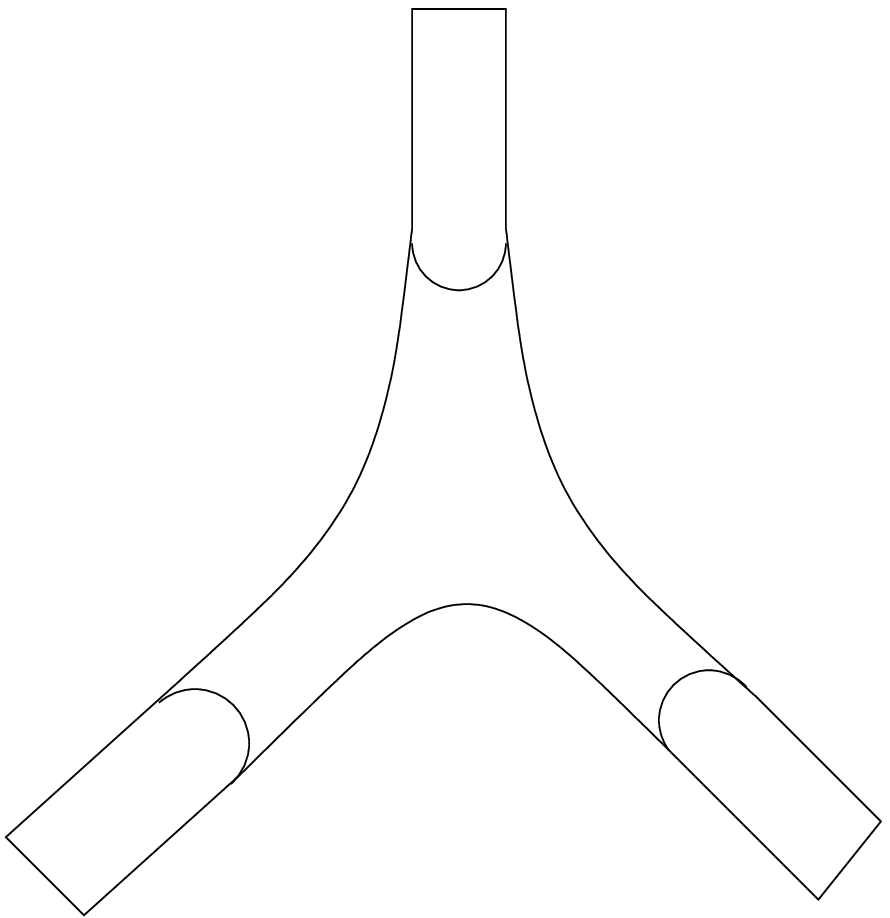}{1.4truein}

There are various possibilities for the form of a $\psi^3$ interaction, 
including possible derivatives on the $\psi$s and the possibility of a 
coupling that varies as some function of the transverse coordinates 
$x_\perp$ (for example, a varying dilaton). Though such interactions do 
seem to give IR singularities with nontrivial dependence on external 
momenta, we have not found a simple action with local interactions of 
the $\psi$s in the bulk that reproduces the two loop singularities 
above.

A second possibility would be that the momentum dependence of the two 
loop IR singularities arises from a loop of $\chi$s (or $\psi$s), such 
as the diagram of figure 5b. The form of the expressions \symb\ and 
\asyma\ are suggestive of such a diagram, if we reinterpret the $y$ 
integrals as integrals over a loop momentum $l = \theta y$. For example, 
the expression \symb\ becomes
\eqn\symbl{
V_1(p_1,p_2,p_3) = {g^5 \pi^3 \over 3 \cdot 2^5} \det(\Theta) 
\int^\Lambda {d^6 l 
\over (2 \pi)^6} {1 \over l \circ l \; (l + p_2) \circ (l +p_2) \; (l-  
p_3) \circ (l - p_3)} \; .}
which is exactly reproduce by the diagram of figure 5b with a coupling 
proportional to $g^{5 \over 3} \det {}^{1 \over 3}(\Theta) \phi(x) 
\chi_0(x) \chi_0(x)$ where $\chi_0$ propagates on the brane with metric 
$g_{\mu \nu}$. Such a coupling is undesirable, however, since it would 
lead to higher point functions with fractional powers of the coupling 
and more severe IR singularities which are not present in the original 
theory.

To conclude, we do not have a satisfactory interpretation of the two 
loop IR singularities in terms of weakly coupled light degrees of 
freedom. We hope to return to this important problem in the near future.

\bigskip
\centerline{\bf Acknowledgements}

We would like to thank T.  Banks, D.  Gross, I.  Klebanov, S.  Minwalla,
G.  Semenoff, L.  Susskind, and E.  Witten for useful discussions.  The
work of M.V.R.  was supported in part by NSF grant PHY98-02484 and that
of N.S.  by DOE grant \#DE-FG02-90ER40542.

\listrefs

\end